\documentclass[conference]{IEEEtran}
\IEEEoverridecommandlockouts
\usepackage{cite}
\usepackage{amsmath,amssymb,amsfonts}
\usepackage{float}
\usepackage{algorithmic}
\usepackage{graphicx}
\usepackage{multirow}
\usepackage{textcomp}
\usepackage{xcolor}
\usepackage{stfloats}
\usepackage{tikz}
\usepackage{bm}
\usepackage{pgfplots}
\usepackage{pgfplotstable}
\pgfplotsset{compat=1.7}
\usepackage{subcaption}
\newtheorem{remark}{Remark}
\newtheorem{lemma}{Lemma}
\def\BibTeX{{\rm B\kern-.05em{\sc i\kern-.025em b}\kern-.08em
    T\kern-.1667em\lower.7ex\hbox{E}\kern-.125emX}}

\begin{document}

\title{Reliability and Latency Analysis of Sliding Network Coding With Re-Transmission
}

\author{\IEEEauthorblockN{Fangzhou Wu\IEEEauthorrefmark{1},
		Zhiyuan Tan\IEEEauthorrefmark{1}, Huiying Zhu\IEEEauthorrefmark{1}, and
		Pengpeng Dong\IEEEauthorrefmark{1}}
	\IEEEauthorblockA{\IEEEauthorrefmark{1}Wireless Network RAN Research Department,
		Huawei Technologies Co., Ltd., Shanghai, China}
	\IEEEauthorblockA{\{fangzhou.wu, tanzhiyuan, zhuhuiying1, d47252\}@huawei.com}
}


\maketitle

\begin{abstract}
Future networks are expected to support various ultra-reliable low-latency communications via wireless links. To avoid the loss of packets and keep the low latency, sliding network coding (SNC) is an emerging technology by generating redundant packets that are the linear combination of the original data packets from the current block and some previous blocks. However, how to take the advantage of re-transmission for SNC is still an open problem since higher reliability could be achieved at the expense of large latency caused by round-trip time (RTT). To deal with this issue, in this paper, we consider the idea of adjusting the transmission phase and the number of the redundant packets for SNC with re-transmission. Specifically, If RTT is large, most of the redundant packets are sent at the first transmission, otherwise, re-transmission will be used. We first derive a concise and tight lower bound of the block error probability of SNC without re-transmission. Then, based on the bound, the theoretical expressions of the proposed re-transmission schemes are derived regarding the block error probability, the average code length, and the average packet latency. Results show that the proposed SNC with re-transmission improves block error probability and keeps the low latency.
\end{abstract}


\section{Introduction}
Network coding (NC) is one of the most promising solutions to enabling ultra-reliable low-latency communications (URLLC) for future networks especially when the wireless link endures channel fading, interference, and burst error. It generates redundant packets at the first transmission via the packet-wise coding to protect the original data packets from loss, which meets low latency requirements.

The notion of NC was first introduced in the multicast network for efficient routing \cite{ahlswede2000network}, which achieves the capacity. Following the analysis on coding theory, \cite{cai2002network} provided the idea of packet-wise coding to correct the lost data packets by transmitting the coded (or superposed) packets instead of the individual original data packets. And the corresponding bounds were derived in \cite{yeung2006network,cai2006network} from an information-theoretic perspective. To approach the given performance bounds, several coding schemes have been proposed, such as the fountain codes \cite{shokrollahi2006raptor}, random linear network coding \cite{ho2006random}, and the batched sparse code \cite{yang2014batched}. The code length (i.e., the number of coded packets within each block) cannot be too long to keep the low latency. This leads to a performance loss on the reliability since the increase in the code length can provide a lower packet loss probability.

To deal with the raised challenge, recent advances showed that the use of the sliding window (or the conventional kernel) can make the designed code not only keep a lower latency compared with the one of the code with a long length but also provide higher reliability compared with the one of the code with a short length. Codes with such a structure are called sliding NC (SNC) \cite{choi2021sliding} or convolutional NC \cite{li2006convolutional,guo2021convolutional}. Unlike the aforementioned block codes, SNC generates the coded packets that are the linear combination of the original data packets from the current block and some previous blocks. Through some careful designs, the length of each block can be short to guarantee the low latency of decoding each packet while the long length of the sliding window (also known as the memory length of convolutional codes) keeps the low packet loss probability. This implies that SNC is a suitable solution to URLLC.

Existing works of SNC mainly concentrated on the case without re-transmission. However, with the use of re-transmission, SNC can achieve higher reliability without increasing the number of redundant packets. The main concern of enabling re-transmission is the latency since it has to consider the influence of the round-trip time (RTT), which may result in an unguaranteed latency. Thus, to keep the low latency and to take the advantage of re-transmission, how to design the re-transmission schemes for SNC needs to be addressed.

Motivated by the above observations, in this paper, we consider SNC with re-transmission and propose three re-transmission schemes by adjusting the transmission phase and the number of the redundant packets, i.e., determining whether the redundant packets are sent at the first transmission, the re-transmission, or both. To this end,  an insightful lower bound of SNC without re-transmission is derived at first. Based on the concise bound, we further derive the theoretical expressions of the block error probability, the average code length, and the average packet latency for the considered re-transmission schemes. In the different use cases with the different RTT, we then can choose the corresponding re-transmission scheme to ensure the latency.

The rest of this paper is arranged as follows. Section \ref{ss2} provides the network model and the proposed re-transmission schemes. In Section \ref{s3}, we first derive an upper bound of the error probability to make the analysis more tractable. Then, three re-transmission schemes are analyzed based on the derived bound. Simulation and comparison among block codes and SNC are provided in Section \ref{s4} and conclusions are given in Section \ref{s5}.

\section{Network Model and Re-Transmission Schemes for RS-SNC}\label{ss2}
In this section, the considered end-to-end network model is described at the beginning. Then, we provide three re-transmission schemes for Reed-Solomon-based SNC (RS-SNC) by adjusting the transmission phase and the number of added redundant packets.

\subsection{Network Model and RS-SNC Basic}
Following \cite{choi2021sliding}, we consider the end-to-end network where the source node wants to send the data packets to the sink node. The binary packet-erasure channel is considered with the erasure probability $ \epsilon $, i.e., the probability of losing $ X=r $ packets among $ n $ packets is
\begin{equation}\label{key}
	\Pr(X=r)= {\tbinom{n}{r}} (1-\epsilon)^{n-r}\epsilon^r=f(r;n,\epsilon),
\end{equation}
where $ X $ follows the binomial distribution.

To ensure the reliability of transmitting packets, RS-SNC is applied in this network, which can be regarded as a $ (n,k,L) $ convolutional code with an infinite generation matrix 
\begin{equation}\label{key}
	\bm{\mathrm{G}}=\begin{bmatrix}
		\ddots&  &  &  \\
		\vdots & \bm{\mathrm{G}}_L &  &  \\
		\ddots & \vdots  & \bm{\mathrm{G}}_L  &  \\
		 & \bm{\mathrm{G}}_0 & \vdots  & \ddots  \\
		 &  & \bm{\mathrm{G}}_0  & \vdots \\
		 &  &  & \ddots
	\end{bmatrix}\stackrel{(a)}{=}\begin{bmatrix}
	\ddots&  &  &  \\
	\vdots & \bm{0}\bm{\mathrm{P}}_L &  &  \\
	\ddots & \vdots  & \bm{0}\bm{\mathrm{P}}_L  &  \\
	& \bm{\mathrm{I}}_k\bm{\mathrm{P}}_0 & \vdots  & \ddots  \\
	&  & \bm{\mathrm{I}}_k\bm{\mathrm{P}}_0  & \vdots \\
	&  &  & \ddots
\end{bmatrix},
\end{equation}
where $ L $ is the memory length of the code (i.e., the length of the sliding window is $ (L+1)n $), $ \bm{\mathrm{G}}_l, \forall l\in[0:L] $, is the generator  matrix of a $ (n,k) $ RS-BC such that $ \bm{\mathrm{G}} $ achieves the maximum distance profile (MDP) \cite{hutchinson2005convolutional} by the careful design \footnote{To show the existence, a straightforward way is that each $ \bm{\mathrm{G}}_l $ is selected without overlap from the same Vandermonde matrix.}\cite{smarandache2001constructions,alfarano2020weighted}, and a systematic RS-SNC is constructed by making $  \bm{\mathrm{G}}_l=[\bm{0}\bm{\mathrm{P}}_l]$ for $ l\in[1:L] $ and $  \bm{\mathrm{G}}_0=[\bm{\mathrm{I}}_k\bm{\mathrm{P}}_0] $ given as the condition $ (a) $. Then, for the MDP RS-SNC, in a sliding window of length $ (l+1)n $, at most $ (l+1)(n-k) $ erasures can be corrected in the erasure channel \cite{tomas2012decoding}.

\subsection{Re-Transmission Schemes}
\begin{figure}
	\centering
	\includegraphics[width=\linewidth]{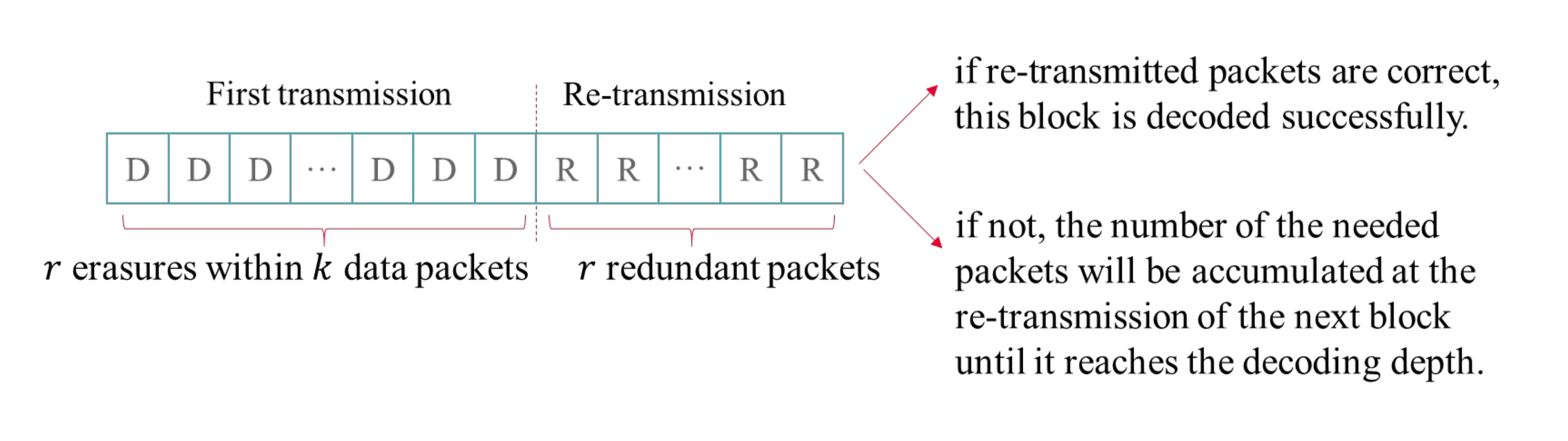}
	\caption{Mode 1.}
	\label{fig:m1}
\end{figure}

We begin with the naive re-transmission scheme as Mode 1 (M1). As shown in Fig.~\ref{fig:m1}, for each block, the first transmission only contains $ k $ original data packets and re-transmits the demanded number of redundant packets depending on how many original data packets are erased. Assume the maximum re-transmission time for each block is $ 1 $.

\begin{figure}
	\centering
	\includegraphics[width=\linewidth]{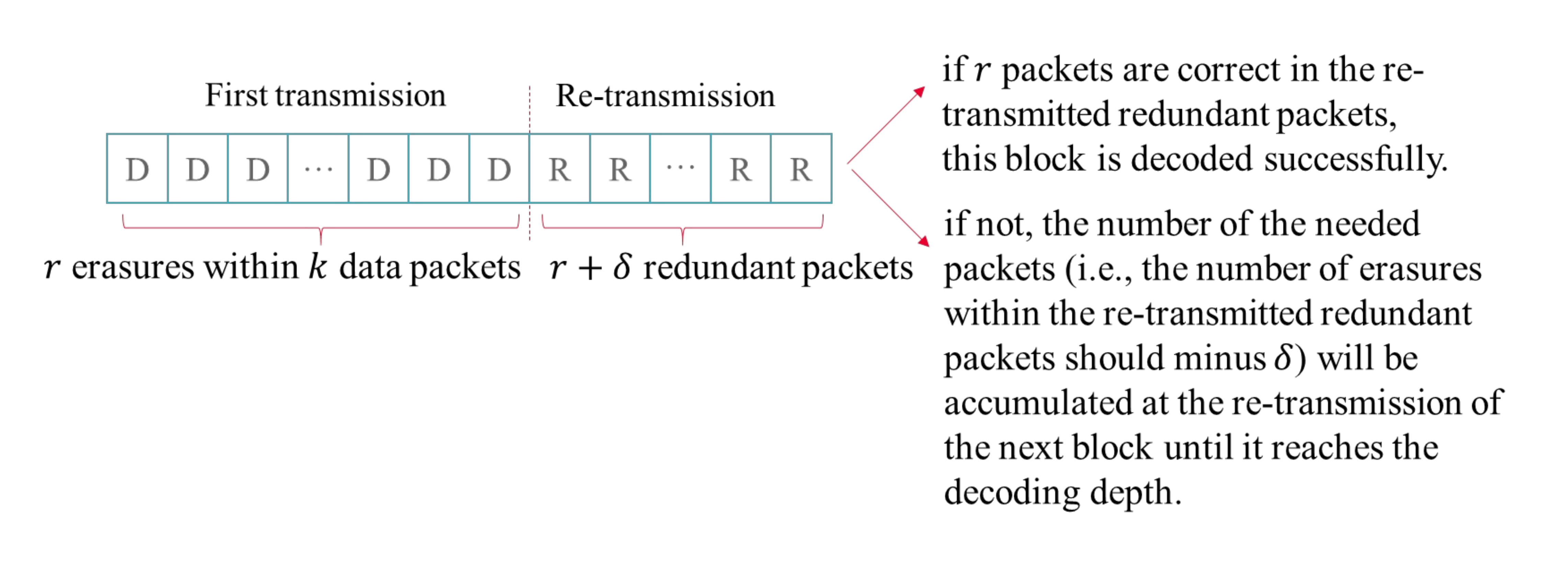}
	\caption{Mode 2.}
	\label{fig:m2}
\end{figure}

In Mode 2 (M2) shown as Fig.~\ref{fig:m2}, the first transmission of each block is the same as the one mentioned in M1 while at the re-transmission phase, if $ r $ packets are lost, $ r+\delta $ redundant packets will be sent where $ \delta $ is the extra redundant packets. Assume $ Y $ follows $ Y\sim B(X+\delta,\epsilon) $.

\begin{figure}
	\centering
	\includegraphics[width=\linewidth]{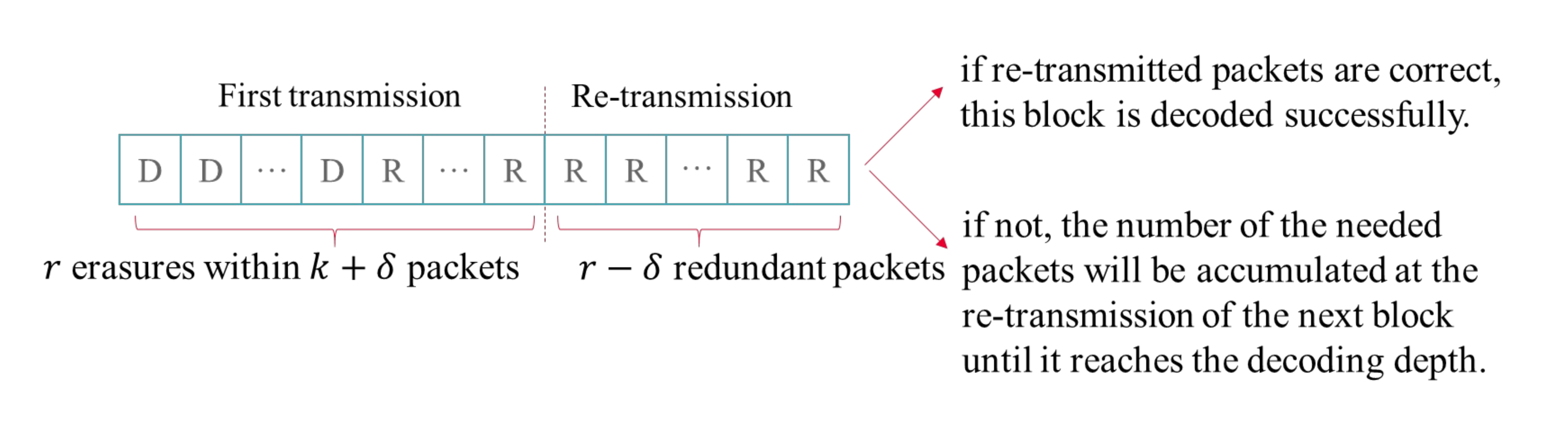}
	\caption{Mode 3.}
	\label{fig:m3}
\end{figure}

Unlike M2 adding the extra redundant packets in the re-transmission phase, Mode 3 (M3) chooses to add the extra $ \delta $ redundant packets in the first transmission instead. 

Based on the analysis in the following section, we will show how RS-SNC with re-transmission achieves an improved success probability while keeping the low packet latency.

\section{Re-Transmission Schemes for RS-SNC}\label{s3}
In this section, we first bound the performance of RS-SNC and compare it with Reed-Solomon block code (RS-BC). Then, based on this lower bound, the analysis of these three re-transmission schemes is provided in terms of the corresponding success probability, average code rate, and average packet latency, respectively.

For RS-SNC, the performance is evaluated by the first-block success probability, the packet latency, and the code rate. As follows, we begin with the analysis of the basic RS-SNC.

\subsection{Analysis of RS-SNC}\label{S2}
\begin{figure*}[b]
	\begin{equation}\label{RS-SNC success probability}
		\begin{split}
			\Pr{\rm suc}=&\underbrace{\sum_{\delta_0=0}^{n-k}f(\delta_0;n,\epsilon)}_{{\rm the\ 1st\ term}}+\underbrace{\sum_{\delta_0=n-k+1}^{n}\sum_{\delta_1=0}^{2(n-k)-\delta_0}f(\delta_0;n,\epsilon)f(\delta_1;n,\epsilon)}_{{\rm the\ 2nd\ term}}\\
			&+\cdots+\underbrace{\sum_{\delta_0=n-k+1}^{n}\cdots\sum_{\delta_{L-1}=L(n-k)-\sum_{j=0}^{L-2}\delta_j+1}^{n}\sum_{\delta_L=0}^{(L+1)(n-k)-\sum_{j=0}^{L-1}\delta_j}f(\delta_0;n,\epsilon)\cdots f(\delta_L;n,\epsilon)}_{{\rm the\ }(L+1){\rm -th\ term}}.
		\end{split}
	\end{equation}
\end{figure*}

\emph{First-Block Success/Error Probability.} Since RS-SNC is not a block code, the performance of RS-SNC is evaluated by the first-block success probability (also known as first-event success/error probability), which is widely used for the analysis of the conventional codes\cite{pietrobon1996probability},\cite[Chapter 4]{johannesson2015fundamentals}. The first-block success probability is defined as the probability of successfully decoding the current block under the condition that the previous blocks are correct.

Suppose that the maximum window size of RS-SNC is $ (L+1)n $, i.e., to decode the $ i $-th block of length $ n $, at most the next $ L $ blocks will be involved. The success probability of decoding the $ i $-th block of length $ n $ is Eq.~\eqref{RS-SNC success probability}.

For $ L=0 $, the above RS-SNC is actually the $ (n,k) $ RS-BC with the success probability\footnote{For block codes, the first-block success/error probability is the same as the block success/error probability.}
\begin{equation}\label{key}
	\Pr{\rm suc}|_{L=0}=\sum_{\delta_0=0}^{n-k}f(\delta_0;n,\epsilon)=F(n-k;n,\epsilon)
\end{equation}
which is seen as a naive lower bound of $ \Pr{\rm suc} $. This implies that the performance of the $ (n,k,L) $ RS-SNC is generally better than the one of the $ (n,k) $ RS-BC.

We also derive a more accurate lower bound in the following lemma, which not only keeps concise but also shows the influence of the memory length $ L $. 
\begin{lemma}\label{lemma lower bound}
	The lower bound of $ \Pr{\rm suc} $ (Eq.~\eqref{RS-SNC success probability}) is given as 
	\begin{equation}\label{lower bound}
		\begin{split}
			\Pr{\rm suc}\ge&\sum_{\delta_0=0}^{n-k}f(\delta_0;n,\epsilon)+\\
			&\sum_{\delta_0=n-k+1}^{n}f(\delta_0;n,\epsilon)F((L+1)(n-k)-\delta_0;Ln,\epsilon).
		\end{split}
	\end{equation}
	\begin{IEEEproof}
		To this end, consider the lower bound of the summation of the 2nd and 3rd terms in Eq.~\eqref{RS-SNC success probability} as an example to show the key idea as follows. The 2nd term in Eq.~\eqref{RS-SNC success probability} is bounded as
		\begin{equation}\label{key relaxion}
			\begin{split}
				&\sum_{\delta_1=0}^{2(n-k)-\delta_0}f(\delta_1;n,\epsilon)\\
				&\ge\sum_{\delta_1=0}^{2(n-k)-\delta_0}\sum_{\delta_2=0}^{3(n-k)-\delta_0-\delta_1}f(\delta_1;n,\epsilon)f(\delta_2;n,\epsilon)
			\end{split}
		\end{equation}
		at first, where $ \sum_{\delta_0=n-k+1}^{n}f(\delta_0;n,\epsilon) $ in this term is omitted temporarily for the sake of simplicity. This makes it possible to be combined with the 3rd term, and the summation is given as
		\begin{equation}
			\begin{split}
				&\sum_{\delta_1=0}^{3(n-k)-\delta_0}\sum_{\delta_2=0}^{3(n-k)-\delta_0-\delta_1}\tbinom{n}{\delta_1}\tbinom{n}{\delta_2}\epsilon^{\delta_1+\delta_2}(1-\epsilon)^{2n-(\delta_1+\delta_2)}\label{terms of 1 and 2}\\
				&\stackrel{(a)}{=}\sum_{\delta=0}^{^{3(n-k)-\delta_0}}\sum_{\delta_1+\delta_2=\delta}\tbinom{n}{\delta_1}\tbinom{n}{\delta_2}\epsilon^{\delta}(1-\epsilon)^{2n-\delta}\\
				&\stackrel{(b)}{=}\sum_{\delta=0}^{^{3(n-k)-\delta_0}}\tbinom{2n}{\delta}\epsilon^{\delta}(1-\epsilon)^{2n-\delta}\\
				&=F(3(n-k)-\delta_0;2n,\epsilon),
			\end{split}
		\end{equation}
		where $ \sum_{\delta_0=n-k+1}^{n}f(\delta_0;n,\epsilon) $ in this term is also omitted temporarily, the condition $ (a) $ follows by changing the upper and lower limits of the double summation, the condition $ (b) $ follows because of Vandermonde's identity.
		
		For the $ 4 $-th term in Eq.~\eqref{RS-SNC success probability}, by using again the idea in Eq.~\eqref{key relaxion}, the summation of the $ 4 $-th term and Eq.~\eqref{terms of 1 and 2} is given as
		$ \sum_{\delta_0=n-k+1}^{n}f(\delta_0;n,\epsilon)F(4(n-k)-\delta_0;3n,\epsilon) $. Using the above steps, Eq.~\eqref{lower bound} is finally obtained.
	\end{IEEEproof}
\end{lemma}


\begin{remark}[Lower Bound]
	There is an intuitive understanding on Eq.~\eqref{lower bound}. If the $ i $-th block of RS-SNC can be decoded successfully with the help of the next block, then whether the rest of the $ L+1 $ blocks is erased or not has no effect on decoding the $ i $-th block. However, for the lower bound, it strictly determines the erasure pattern of each packet of all the next $ L $ blocks, although some blocks may be unrelated to the decoding of the $ i $-th block in some cases. This results in fewer possible combinations and is a lower bound.
\end{remark}

\emph{Average Packet Latency.} We consider the case that the RS-SNC is a systematic code\footnote{This case is widely used in the real communication systems since the systematic code can always provide a lower packet latency than the non-systematic one.}. If the $ p $-th data packet in the $ i $-th block is received correctly, then it can be decoded directly due to the characteristic of the systematic code. If not, the $ p $-th data packet cannot be decoded unless the whole block with $ n $ packets is received. Since at most the next $ L $ blocks can help the decoding of the current block, the distribution of the packet latency of the $ p $-th data packet is given as

\begin{equation}\label{packet latency}
	\begin{split}
				\Pr(D^{\rm SNC}_{p}=d)=\left\lbrace\begin{array}{ll}
				1-\epsilon,	&d=0  \\
			\multirow[t]{2}{*}{$\epsilon\varDelta_l, $} &d=(l+1)n-p\\
			& {\rm for\ }l\in[0:L-1]\\
			\epsilon(1-\sum_{l=0}^{L-1}\varDelta_l),&  (L+1)n-p\\
			0, &{\rm otherwise}
		\end{array} \right.,
	\end{split}
\end{equation}
where $ \varDelta_l $ is the $ (l+1) $-th term in Eq.~\eqref{RS-SNC success probability}.

One can observe that the average packet latency $ \mathsf{E}\left[\sum_{p=1}^{k}D^{\rm SNC}_p/k \right]  $ of the $ (n,k,L) $ RS-SNC is between the one of the $ (n,k) $ RS-BC with the distribution 
\begin{equation}\label{short code}
	\begin{split}
		\Pr(D^{\rm BC}_{p}=d)=\left\lbrace\begin{array}{ll}
			1-\epsilon,	&d=0 \\
			\epsilon, & d=n-p\\
			0, &{\rm otherwise}
		\end{array} \right.
	\end{split}
\end{equation}
and the one of the $ ((L+1)n,(L+1)k) $ RS-BC with the distribution
\begin{equation}\label{long code}
	\begin{split}
		\Pr(D^{\rm BC}_{p}=d)=\left\lbrace\begin{array}{ll}
			1-\epsilon,	& d=0 \\
			\epsilon, & d=(L+1)n-p\\
			0, & {\rm otherwise}
		\end{array} \right..
	\end{split}
\end{equation}

\begin{remark}[Comparable RS-BC]
	We compare the $ (n,k,L) $ RS-SNC with the $ (n,k) $ RS-BC and the $ ((L+1)n,(L+1)k) $ RS-BC in terms of the success probability and the average packet latency. From Eq.~\eqref{lower bound}, it is obvious that the $ (n,k,L) $ RS-SNC can provide better performance than the short RS-BC. Besides, we also show $ \Pr{\rm suc}\ge F((L+1)(n-k);(L+1)n,\epsilon) $ by using Eq.\eqref{key relaxion} to bound the 1st term in Eq.\eqref{lower bound}. This implies that the $ (n,k,L) $ RS-SNC is also comparable with the long RS-BC. As for the average packet latency, in the ultra-reliable scenario where the success probability needs to be guaranteed as $ 99.999\% $ or higher, $ \varDelta_0 $ and $ \varDelta_1 $ are the leading order which contributes the most part of $ \Pr{\rm suc} $. This leads that the maximum packet latency is normally not greater than  $ 2n-p $.
\end{remark}

Thus, RS-SNC is a suitable solution to URLLC since the success probability outperforms the one of the $ ((L+1)n,(L+1)k) $ RS-BC and the packet latency keeps lower. This implies that RS-SNC not only achieves the reliability of the long code but also keeps the low latency provided by the short code.

\subsection{Analysis of Proposed Re-Transmission Schemes}

As shown in Section \ref{S2}, it is difficult to analyze the success probability (Eq.\eqref{RS-SNC success probability}) of decoding the $ i $-th block directly. Thus, to make the problem more tractable, we consider the lower bound of the success probability instead. Clearly, the lower bound (Eq.~\eqref{lower bound}) consists of two parts, the success probability $ \Pr{\rm suc}|_{L=0} $ of directly decoding the $ i $-th block and the one of decoding the $ i $-th block involving the next $ L $ blocks, both of which are heavily related to the probability $ \Pr{\rm suc}|_{L=0} $. This implies that we should focus on the analysis of $ \Pr{\rm suc}|_{L=0} $.

\emph{Mode 1}. The probability of directly decoding the $ i $-th block is given as 
\begin{equation}\label{m1 Pr}
	\begin{split}
		\Pr{\rm suc}|_{L=0}&=\Pr(X=0)+\sum_{r=1}^{k}(1-\epsilon)^{r}\Pr(X=r)\\
		&=(1-\epsilon^2)^k,
	\end{split}
\end{equation}
where $ X $ is the binomial distribution of $ X\sim B(k,\epsilon) $ and $ Y $ follows the binomial distribution of $ Y\sim B(X,\epsilon) $. Besides, due to the consideration of re-transmission, the code rate (length) is dynamic. Thus, we derive the average code length to characterize its performance, which is given as
\begin{equation}\label{key}
	\begin{split}
		n_{\rm M1}&=k\Pr(X=0)+\sum_{r=1}^{k}(k+r)\Pr(X=r)=k+\epsilon k.
	\end{split}
\end{equation}
The distribution of the packet latency of the $ p $-th data packet in Mode 1 is
\begin{equation}\label{packet latency of m1}
	\begin{split}
		\Pr(D^{\rm M1}_{p}=d)=\left\lbrace\begin{array}{ll}
			1-\epsilon, & d=0 \\
			  \multirow[t]{2}{*}{$\epsilon f(r;k-1,\epsilon),$}&d=k+r\\
			  &\quad +1+N_{\rm Re}-p\\
			  0,&{\rm otherwise}
		\end{array}\right.
	\end{split}
\end{equation}
for each $ r\in[0:k-1] $, where $ N_{\rm Re} $ stands for the latency related to RTT.

\emph{Mode 2.} $ \Pr{\rm suc}|_{L=0} $ is given as 
\begin{equation}\label{M2 Pr}
	\begin{split}
		\Pr{\rm suc}|_{L=0}&=\sum_{r=0}^{k}\Pr(Y\le\delta|X=r)\Pr(X=r)\\
		&=\sum_{r=0}^{k}f(r;k,\epsilon)F(\delta;r+\delta,\epsilon),
	\end{split}
\end{equation}
the average code length is 
\begin{equation}\label{M2 Code}
	\begin{split}
		n_{\rm M2}&=\sum_{r=0}^k(k+r+\delta)\Pr(X=r)-\delta\Pr(X=0)\\
		&=k+\delta+\epsilon k-\delta(1-\epsilon)^k,
	\end{split}
\end{equation}
and the distribution of the packet latency of the $ p $-th data packet in Mode 2 is
\begin{equation}\label{packet latency of m2}
	\begin{split}
		\Pr(D^{\rm M2}_{p}=d)=\left\lbrace\begin{array}{ll}
			1-\epsilon,&d=0 \\
			 \multirow[t]{2}{*}{$ \epsilon f(r;k-1,\epsilon), $}&d=k+\delta+r\\
			&\quad +1+N_{\rm Re}-p\\
			0,&{\rm otherwise}
		\end{array} \right.
	\end{split}
\end{equation}
for each $ r\in[0:k-1] $.

\emph{Mode 3.} We have 
\begin{equation}\label{m3 pr}
	\begin{split}
		\Pr{\rm suc|_{L=0}}=&\sum_{r=0}^{\delta}\Pr(X=r)+\sum_{r=\delta+1}^{k+\delta}(1-\epsilon)^{r-\delta}\Pr(X=r)\\
		=&F(\delta;k+\delta,\epsilon)\\
		&+(1-\epsilon)^k(1+\epsilon)^{k+\delta}F\left(k-1;k+\delta,\hat{\epsilon}\right)
	\end{split}		
\end{equation}
by introducing $ \hat{\epsilon}=\frac{1}{1+\epsilon} $,
\begin{equation}\label{key}
	\begin{split}
		n_{\rm M3}&=\sum_{r=0}^{\delta}(k+\delta)\Pr(X=r)+\sum_{r=\delta+1}^{k+\delta}(r+k)\Pr(X=r)\\
		&=\sum_{r=0}^{\delta}(\delta-r)f(r;k+\delta,\epsilon)+k+\epsilon (k+\delta),
	\end{split}
\end{equation}
and
\begin{equation}\label{packet latency of m3}
	\begin{split}
		\Pr(D^{\rm M3}_{p}=d)=\left\lbrace\begin{array}{ll}
			1-\epsilon,	& d=0 \\
			\multirow[t]{2}{*}{$\epsilon f(r;k-1,\epsilon)  $,} &d=k+\delta-p\\
			& {\rm for\ } r\in[0:\delta-1]\\
			\multirow[t]{3}{*}{$ \epsilon f(r;k-1,\epsilon), $} &d=k+r+1\\
			&\quad +N_{\rm Re}-p-\delta\\ 
			& {\rm for\ } r\in[\delta:k-1]\\
			0,&{\rm otherwise}
		\end{array} \right.,
	\end{split}
\end{equation}
accordingly.

\begin{remark}[Characteristics of Three Re-Transmission Modes]
	Some characteristics of the above three modes are given as follows based on the derived results. 
	\begin{itemize}
		\item Mode 1: In this mode, the average code rate is fixed as $ 1/(1+\epsilon) $, which cannot be flexibly adjusted. And the success probability decreases rapidly with the increase in $ k $. This implies that, for URLLC considering the reliability of $ 99.999\% $ or higher, the naive re-transmission mode is not suitable. 
		\item Mode 2 and Mode 3: A way to improve the reliability is to add some extra redundant packets in either the first transmission (as M3) or the re-transmission (as M2). This would ensure that the success probability satisfies the target. By comparing $ n_{\rm M2} $ with $ n_{\rm M3} $, one can observe that the throughput of M2 is greater than the one of M3 if $ \epsilon $ is small, but the latency would become large if RTT increases. This means we need to switch modes with respect to the different cases to meet the guaranteed reliability and latency.
	\end{itemize}
\end{remark}

\begin{remark}[Special Cases]
	When $ \delta $ is equal to 0, the success probability of M2 and M3 is reduced to the same one of M1, which verifies the correctness of the derivation.
	\begin{itemize}
		\item Mode 2. By setting $ \delta=0 $ in $ \Pr{\rm suc}|_{L=0} $, we have $ \Pr{\rm suc}|_{L=0}=F(0;k,\epsilon^2)=(1-\epsilon^2)^k $.
		\item Mode 3. By setting $ \delta=0 $ in $ \Pr{\rm suc}|_{L=0} $, we have $ \Pr{\rm suc}|_{L=0}=F(0;k,\epsilon)+(1-\epsilon)^k(1+\epsilon)^{k}F\left(k-1;k,{1}/{1+\epsilon}\right)=(1-\epsilon)^k+(1-\epsilon)^k(1+\epsilon)^k(1-(1+\epsilon)^{-k})=(1-\epsilon)^k(1+\epsilon)^k=(1-\epsilon^2)^k$.
	\end{itemize}
\end{remark}

\section{Simulation Results}\label{s4}
In this section, we provide some simulation results to verify the derived bounds, first-block error probability  (i.e, 1-$ \Pr{\rm suc} $), and average packet latency. Further, three re-transmission schemes are compared and their characteristics are demonstrated.

We consider the random erasure channel where the packet erasure probability follows the binomial distribution with the parameter $ \epsilon $. For the re-transmission case, the maximum number of re-transmissions is set to one.

Fig.~\ref{fig:1} shows the comparison of the first block error probability among RS-SNC and the RS block codes. As analyzed in Section \ref{S2}, RS-SNC achieves a lower error probability compared with the corresponding RS block code. In other words, $ (n,k,L) $ RS-SNC outperforms the comparable $ (n(L+1),k(L+1)) $ RS block code. When $ L=1 $, the derived bound is the same as the simulated one, which verifies the correctness of  Eq.~\eqref{lower bound}.

{\begin{figure}[!ht]
	\centering
	\pgfplotstableread[col sep=comma,]{CS1.csv}\datatable
	\begin{tikzpicture}[scale=0.85]
		\begin{axis}[
			width=\linewidth,
			ymode=log,
			ymax=1,
			xmin=0.1,xmax=0.3,
			grid=both,
			x tick label style={font=\normalsize},
			legend style={nodes={scale=0.5, transform shape},at={(1,0.01)},anchor=south east},
			legend cell align={left},
			ylabel={First-Block Error Probability},
			xlabel={Packet Erasure Probability $\epsilon$}]
			
			\addplot [black] table [x={epsilon}, y={k_8_L_1_rate_2_3_L_L}]{\datatable};
			\addlegendentry{$(n(L+1),k(L+1))$ RS-BC};
			
			\addplot [dashed, mark=o, mark options={solid},black] table [x={epsilon}, y={k_8_L_1_rate_2_3_L_0}]{\datatable};
			\addlegendentry{$(n,k)$ RS-BC, $ n=12, k=8 $}
			
			\addplot [dashed, mark=*, mark options={solid},black] table [x={epsilon}, y={k_12_L_1_rate_2_3_L_0}]{\datatable};
			\addlegendentry{$(n,k)$ RS-BC, $ n=18, k=12 $};

			\addplot [mark=o, mark options={solid},red] table [x={epsilon}, y={k_8_L_2_rate_2_3_LB}]{\datatable};
			\addlegendentry{$(n,k,L)$ RS-SNC, the upper bound, $ 1-\Pr{\rm suc} $ (Eq.~\eqref{lower bound})}
			
			\addplot [mark=square, mark options={solid},blue] table [x={epsilon}, y={s_k_8_L_1_rate_2_3_LB}]{\datatable};
			\addlegendentry{$(n,k,L)$ RS-SNC, simulated};
		
			\addplot [black] table [x={epsilon}, y={k_12_L_1_rate_2_3_L_L}]{\datatable};
			
			\addplot [mark=o, mark options={solid},red] table [x={epsilon}, y={k_8_L_1_rate_2_3_LB}]{\datatable};
			
			\addplot [black] table [x={epsilon}, y={k_12_L_2_rate_2_3_L_L}]{\datatable};
			
			\addplot [mark=o, mark options={solid},red] table [x={epsilon}, y={k_12_L_2_rate_2_3_LB}]{\datatable};

			\addplot [mark=square, mark options={solid},blue] table [x={epsilon}, y={s_k_8_L_2_rate_2_3_LB}]{\datatable};

			\addplot [mark=square, mark options={solid},blue] table [x={epsilon}, y={s_k_12_L_2_rate_2_3_LB}]{\datatable};

		\end{axis}
	\draw (0.5,3.3) ellipse (0.1 and 0.3);
	\node (A) at (0.65, 3.3) {};
	\node (B) at (4.5, 3.3) {\small
		 $ n=12, k=8, L=1 $};
	\draw[dashed,->] (A) -- (B);
	\draw (0.5,2.5) ellipse (0.1 and 0.4);
		\node (C) at (0.65, 2.5) {};
	\node (D) at (4.5, 2.5) {\small
		 $ n=12, k=8, L=2 $};
	\draw[dashed,->] (C) -- (D);
	\draw (0.5,1.3) ellipse (0.1 and 0.53);
		\node (E) at (0.65, 1.4) {};
	\node (F) at (4.5, 1.7) {\small
		 $ n=18, k=12, L=2 $};
	\draw[dashed,->] (E) -- (F);
	\end{tikzpicture}
	\caption{First-block error probabilities of RS-BC and RS-SNC with respect to the packet erasure probability and the length of the sliding window.}
	\label{fig:1}
\end{figure}
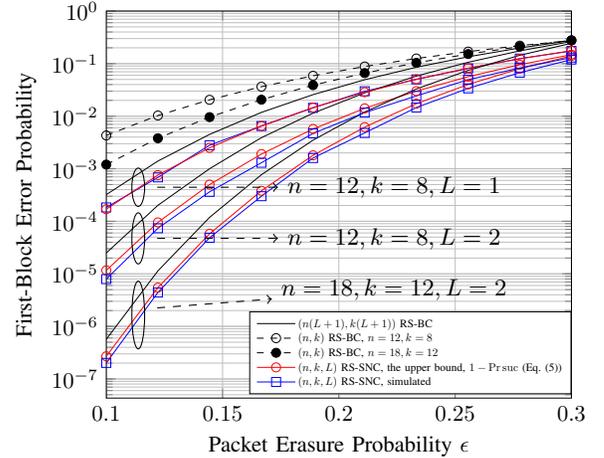}

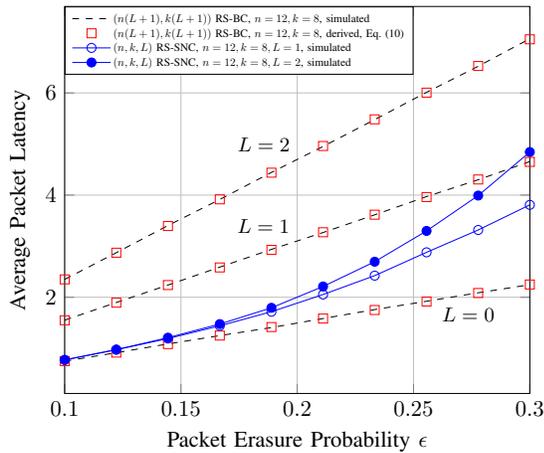
\begin{figure}[!ht]
	\centering
	\pgfplotstableread[col sep=comma,]{CS1D.csv}\datatable
	\begin{tikzpicture}[scale=0.85]
		\begin{axis}[
			width=\linewidth,
			xmin=0.1,xmax=0.3,
			grid=both,
			x tick label style={font=\normalsize},
			legend style={nodes={scale=0.5, transform shape},at={(0,1)},anchor=north west},
			legend cell align={left},
			ylabel={Average Packet Latency},
			xlabel={Packet Erasure Probability $ \epsilon $}]
			
			\addplot [dashed, black] table [x={epsilon}, y={d_k_8_1_L_0_rate_2_3}]{\datatable}node [pos=0.8, below right] {\small$L=0$};
			\addlegendentry{$(n(L+1),k(L+1))$ RS-BC, $ n=12,k=8 $, simulated};
			
			\addplot [only marks, mark=square, mark options={solid},red] table [x={epsilon}, y={d_d_k_8_1_L_0_rate_2_3}]{\datatable};
			\addlegendentry{$(n(L+1),k(L+1))$ RS-BC, $ n=12,k=8 $, derived, Eq.~\eqref{long code}};

			\addplot [mark=o, mark options={solid},blue] table [x={epsilon}, y={d_k_8_1_L_1_rate_2_3}]{\datatable};
			\addlegendentry{$(n,k,L)$ RS-SNC, $ n=12,k=8,L=1 $, simulated}
			
			\addplot [mark=*, mark options={solid},blue] table [x={epsilon}, y={d_k_8_1_L_2_rate_2_3}]{\datatable};
			\addlegendentry{$(n,k,L)$ RS-SNC, $ n=12,k=8,L=2 $, simulated};
			
			\addplot [dashed,black] table [x={epsilon}, y={d_k_8_2_L_0_rate_2_3}]{\datatable}node [pos=0.5, above left] {\small$L=1$};
			
			\addplot [dashed,black] table [x={epsilon}, y={d_k_8_3_L_0_rate_2_3}]{\datatable}node [pos=0.5, above left] {\small$L=2$};
			
			\addplot [only marks, mark=square, mark options={solid},red] table [x={epsilon}, y={d_d_k_8_2_L_0_rate_2_3}]{\datatable};
			
			\addplot [only marks, mark=square, mark options={solid},red] table [x={epsilon}, y={d_d_k_8_3_L_0_rate_2_3}]{\datatable};

		\end{axis}
	\end{tikzpicture}
	\caption{Average packet latency of RS-BC and RS-SNC with respect to the packet erasure probability and the length of the sliding window.}
	\label{fig:2}
\end{figure}

We also present the average packet latency of RS-SNC and RS-BC in Fig.~\ref{fig:2}. Clearly, the latency of the $ (n,k,L) $ RS-SNC is between the one of $ (n(L+1),k(L+1)) $ RS-BC and the one of $ (n,k) $ RS-BC. Combining the results of Fig.~\ref{fig:1} and Fig.~\ref{fig:2}, we can conclude that RS-SNC is more reliable compared with the long code and also keeps low latency close to the one of the short code, especially for URLLC with the success probability of $ 99.999\%$ or higher.

When the re-transmission enables, the block error probability can be further decreased. As shown in Fig.~\ref{fig:3}, under the same average code length, both M2 and M3 can provide improved performance compared with the classical RS-SNC. Especially, M2 outperforms M3,  because, in the re-transmission phase, M2 provides a lower code rate, which leads to higher reliability.

\begin{figure}[!ht]
	\centering
	\pgfplotstableread[col sep=comma,]{CS2D.csv}\datatable
	\begin{tikzpicture}[scale=0.85]
		\begin{axis}[
			width=\linewidth,
			xmin=0.15,xmax=0.3,
			grid=both,
			x tick label style={font=\normalsize},
			legend entries={{Latency of RS-SNC and RS-BC provided in Eqs.~\eqref{long code}, \eqref{packet latency}, and \eqref{short code}, respectively},
				{$ L=0, N_{\rm re}=1 $, derived},
				{$ L=0, N_{\rm re}=1 $, simulated},
				{$ L=1, N_{\rm re}=1 $, simulated},
				{$ L=1, N_{\rm re}=8 $, simulated},
				{($ n_{\rm M1},k,L $)  RS-SNC, $ \lceil n_{\rm M1}\rceil=12,k=8$},
				{($ n_{\rm M2},k,L $)  RS-SNC,$ \lceil n_{\rm M2}\rceil=12,k=8$},
				{($ n_{\rm M3},k,L $)  RS-SNC,$ \lceil n_{\rm M3}\rceil=12,k=8$},
			},
			legend style={nodes={scale=0.5, transform shape},at={(0,1)},anchor=north west},
			legend cell align={left},
			ylabel={Average Packet Latency},
			xlabel={Packet Erasure Probability $ \epsilon $}]
			\addlegendimage{no markers, dotted, blue,line width=1}
			\addlegendimage{only marks,red,mark=o}
			\addlegendimage{no markers,dashed,black}
			\addlegendimage{no markers,black}
			\addlegendimage{no markers,red, dash dot}
			\addlegendimage{only marks, mark=square}
			\addlegendimage{only marks, mark=diamond}
			\addlegendimage{only marks, mark=triangle}

			\addplot [only marks, mark=o, mark options={solid},red,mark repeat=2,mark phase=2] table [x={epsilon}, y={d_re_k_8_m1_ave}]{\datatable};
			\addplot [dashed,mark=square, mark options={solid,black}, black,mark repeat=2,mark phase=0] table [x={epsilon}, y={d_re_k_8_m1_L_0}]{\datatable};
			\addplot [mark=square, mark options={solid,black}, black] table [x={epsilon}, y={d_re_k_8_m1_L_1}]{\datatable};
			\addplot [mark=square, dash dot, mark options={solid,black}, red] table [x={epsilon}, y={d_re_k_8_m1_L_1_8}]{\datatable};
			\addplot [only marks, mark=o, mark options={solid},red,mark repeat=2,mark phase=2] table [x={epsilon}, y={d_re_k_8_m2_ave}]{\datatable};
			\addplot [dashed, mark=diamond, mark options={solid,black}, black,mark repeat=2,mark phase=0] table [x={epsilon}, y={d_re_k_8_m2_L_0}]{\datatable};
			\addplot [mark=diamond, mark options={solid,black}, black] table [x={epsilon}, y={d_re_k_8_m2_L_1}]{\datatable};
			\addplot [mark=diamond,dash dot, mark options={solid,black}, red] table [x={epsilon}, y={d_re_k_8_m2_L_1_8}]{\datatable};
			\addplot [only marks, mark=o, mark options={solid},red,mark repeat=2,mark phase=2] table [x={epsilon}, y={d_re_k_8_m3_ave}]{\datatable};
			\addplot [dashed,mark=triangle, mark options={solid,black}, black,mark repeat=2,mark phase=0] table [x={epsilon}, y={d_re_k_8_m3_L_0}]{\datatable};
			\addplot [mark=triangle, mark options={solid,black}, black] table [x={epsilon}, y={d_re_k_8_m3_L_1}]{\datatable};
			\addplot [mark=triangle, dash dot,mark options={solid,black}, red] table [x={epsilon}, y={d_re_k_8_m3_L_1_8}]{\datatable};
			\addplot [dotted,blue,line width=1] table [x={epsilon}, y={d_re_k_8_L_0}]{\datatable};
			\addplot [dotted, blue,line width=1] table [x={epsilon}, y={d_re_k_8_2_L_0}]{\datatable};
			\addplot [dotted, blue,line width=1] table [x={epsilon}, y={d_re_k_8_L_1}]{\datatable};

		\end{axis}
	\end{tikzpicture}
	\caption{Average packet latency of RS-BC, RS-SNC, and the re-transmission schemes with respect to the packet erasure probability and the length of the sliding window.}
	\label{fig:4}
\end{figure}
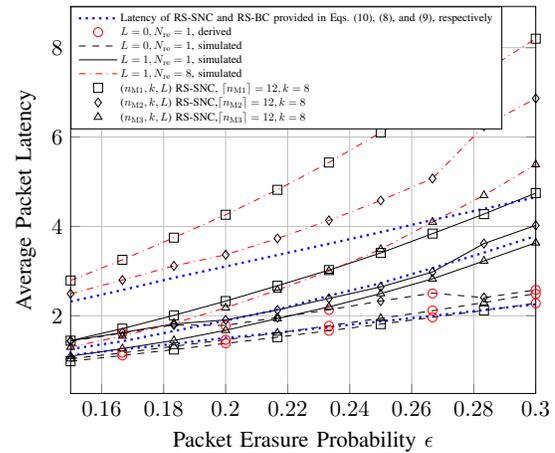

However, in terms of the average packet latency, we can observe that M3 is the lowest one compared with the other two modes in Fig.~\ref{fig:4}. Combining the results of Fig.~\ref{fig:3} and Fig.~\ref{fig:4}, for the large RTT (i.e., $ N_{\rm re}=8 $), M3 is the optimal choice under the small $ \epsilon $. if $ N_{\rm re} $ is small, M2 is the best one of which the average latency is close to M3 and the block error probability is the lowest one. This means that SNC with re-transmission can provide improved reliability and meet the required latency by the switch of the phase that transmits the redundant packets under the same average code length.

\begin{figure}[!ht]
	\centering
	\pgfplotstableread[col sep=comma,]{CS2.csv}\datatable
	\begin{tikzpicture}[scale=0.85]
		\begin{axis}[
			width=\linewidth,
			ymode=log,
			ymax=1,
			xmin=0.15,xmax=0.3,
			grid=both,
			x tick label style={font=\normalsize},
			legend entries={{$ L=0 $, derived, Eq.~\eqref{m1 Pr}, Eq.~\eqref{m3 pr}, and Eq.~\eqref{M2 Pr}},
				{$ L=0 $, simulated},
				{$ L=1 $, simulated},
				{($ n_{\rm M1},k,L $)  RS-SNC, $ \lceil n_{\rm M1}\rceil=12,k=8$},
				{($ n_{\rm M2},k,L $)  RS-SNC,$ \lceil n_{\rm M2}\rceil=12,k=8$},
				{($ n_{\rm M3},k,L $)  RS-SNC,$ \lceil n_{\rm M3}\rceil=12,k=8$},
				{$(n,k,L)$ RS-SNC, $ n=12,k=8,L=1 $}
			},
			legend style={nodes={scale=0.5, transform shape},at={(1,0.01)},anchor=south east},
			legend cell align={left},
			ylabel={First-Block Error Probability},
			xlabel={Packet Erasure Probability $ \epsilon $}]
			\addlegendimage{only marks,red,mark=o}
			\addlegendimage{no markers, dashed, black}
			\addlegendimage{no markers,red}
			\addlegendimage{only marks, mark=square}
			\addlegendimage{only marks, mark=diamond}
			\addlegendimage{only marks, mark=triangle}
			\addlegendimage{no markers, dotted, blue,line width=1}
			
			\addplot [dotted,blue, line width=1] table [x={epsilon}, y={re_k_8_L_1_rate_2_3_LB}]{\datatable};		
			\addplot [only marks, mark=o, mark options={solid},red,mark repeat=2,mark phase=2] table [x={epsilon}, y={re_k_8_m1_fail}]{\datatable};
			\addplot [mark=square, dashed, mark options={solid,black}, black,mark repeat=2,mark phase=0] table [x={epsilon}, y={re_k_8_m1_L_0}]{\datatable};
			\addplot [mark=square, mark options={solid,black},red] table [x={epsilon}, y={re_k_8_m1_L_1}]{\datatable};
			\addplot [only marks, mark=o, mark options={solid},red,mark repeat=2,mark phase=2] table [x={epsilon}, y={re_k_8_m2_fail}]{\datatable};
			\addplot [mark=diamond,  dashed,mark options={solid,black}, black,mark repeat=2,mark phase=0] table [x={epsilon}, y={re_k_8_m2_L_0}]{\datatable};
			\addplot [mark=diamond, mark options={solid,black},red] table [x={epsilon}, y={re_k_8_m2_L_1}]{\datatable};
			\addplot [only marks, mark=o, mark options={solid},red,mark repeat=2,mark phase=2] table [x={epsilon}, y={re_k_8_m3_fail}]{\datatable};
			\addplot [mark=triangle, dashed, mark options={solid,black}, black,mark repeat=2,mark phase=0] table [x={epsilon}, y={re_k_8_m3_L_0}]{\datatable};
			\addplot [mark=triangle, mark options={solid,black},red] table [x={epsilon}, y={re_k_8_m3_L_1}]{\datatable};

		\end{axis}
	\end{tikzpicture}
	\caption{First-block error probabilities of RS-BC, RS-SNC, and the re-transmission schemes with respect to the packet erasure probability and the length of the sliding window.}
	\label{fig:3}
\end{figure}

\section{Conclusions}\label{s5}
In this paper, we have considered three re-transmission schemes to enhance the performance of SNC. Firstly, the concise and tight lower bound has been derived for SNC without re-transmission. Then, with the help of the bound, we have derived the corresponding expressions for the proposed re-transmission schemes. Through adjusting the phase of sending redundant packets, SNC with re-transmission has achieved a lower first-block error probability and kept the low latency under the same average code length compared with the conventional SNC and block codes.

\section*{Acknowledgment}
The authors would like to thank Dr. Yang Shenghao for his
advice and proofreading of our manuscript.

\bibliographystyle{ieeetr}
\bibliography{NOAH.bib} 

\end{document}